\pdfoutput=1
\documentclass[twocolumn,showpacs]{revtex4-1}
\usepackage{graphicx}
\usepackage{amsfonts}
\usepackage{amsmath} % for unnumbered equations
\usepackage{wasysym} % for hexagon

\usepackage[]{hyperref}
\hypersetup{
  colorlinks   = true, %Colours links instead of ugly boxes
  urlcolor     = blue, %Colour for external hyperlinks
  linkcolor    = blue, %Colour of internal links
  citecolor   = red %Colour of citations
}

\newcommand{\ket}[1]{\,|#1 \rangle}
\newcommand{\initlat}{{\cal L}}
\newcommand{\prob}{\mathop{\mathbf{Pr}}\nolimits}
\newcommand{\connected}{\leftrightarrow}
\newcommand{\prAB}{\prob(A\connected B)}
\newcommand{\pckag}{{p^{\text{kag}}_c}}
\newcommand{\pcsq}{{p^{\square}_c}}
\newcommand{\pchex}{{p^{\hexagon}_c}}
\newcommand{\initstate}{\ket{\alpha}}
\newcommand{\outcome}{{\ket{\phi_m}}}
\newcommand{\outcomeset}{\{\outcome\}}
\newcommand{\pset}{\{p_m\}}
\newcommand{\pmin}{p_{\text{min}}}
\newcommand{\pmax}{p_{\text{max}}}
\newcommand{\qpc}{\hat{\alpha}_c}
\newcommand{\qpchi}{\hat{\alpha}^*_c}

\newcommand{\arrowU}{\negmedspace\uparrow}
\newcommand{\arrowD}{\negmedspace\downarrow}
\newcommand{\arrowUU}{\negmedspace\uparrow\uparrow}
\newcommand{\arrowUD}{\negmedspace\uparrow\downarrow}
\newcommand{\arrowDU}{\negmedspace\downarrow\uparrow}
\newcommand{\arrowDD}{\negmedspace\downarrow\downarrow}
\newcommand{\id}{\openone}
\newcommand{\introword}[1]{\textbf{#1}}

\newcommand{\figref}[1]{Fig.~\ref{#1}}
\newcommand{\condone}{\textit{i}}
\newcommand{\condtwo}{\textit{ii}}

\newcommand{\mysubsec}[1]{\subsection{#1}}

\begin{document}
\title{The role of local and global geometry in quantum entanglement percolation}

\author{Gerald John \surname{Lapeyre} Jr.}
\affiliation{ICFO--Institut de Ci\`encies Fot\`oniques, Mediterranean
Technology Park, 08860 Castelldefels, Spain}

\date{\today}

\begin{abstract}
  We prove that enhanced entanglement percolation via lattice
  transformation is possible even if the new lattice is more poorly
  connected in that: \condone) the coordination number (a local
  property) decreases, or \condtwo) the classical percolation
  threshold (a global property) increases.  In searching for protocols
  to transport entanglement across a network, it seems reasonable to
  try transformations that increase connectivity.  In fact, all
  examples that we are aware of violate both conditions \condone\ and
  \condtwo. One might therefore conjecture that all good
  transformations must violate them. Here we provide a counter-example
  that satisfies both conditions by introducing a new method, partial
  entanglement swapping. This result shows that a transformation may
  not be rejected on the basis of satisfying conditions \condone\ or
  \condtwo. Both the result and the new method constitute
  steps toward answering basic questions, such as whether there is a
  minimum amount of local entanglement required to achieve long-range
  entanglement.
\end{abstract}

%% 03.67.Bg Entanglement production and manipulation (for entanglement
%%  in Bose-Einstein condensates, see 03.75.Gg)

%% 03.67.Hk 	Quantum communication

%% 64.60.ah 	Percolation

%% 03.67.Pp Quantum error correction and other methods for protection
%%  against decoherence (see also 03.65.Yz Decoherence; open systems;
%%  quantum statistical methods

%%  03.67.Mn Entanglement measures, witnesses, and other
%%  characterizations (see also 03.65.Ud Entanglement and quantum
%%  nonlocality


\maketitle

\section{Introduction}

Distribution of quantum entanglement on networks has been studied
vigorously over the past few years. This has been driven by the fact
that entanglement is the fundamental resource in quantum information,
but it is created locally via interaction, while it is often consumed
in systems with widely separated components.  In an ideal
description, each node of the network represents a collection of
qubits, and each edge or link represents entangled states of qubits in
different nodes.

But, even in the case of transporting entanglement along a chain of
partially entangled pure states, using perfect quantum
operations, the resulting entanglement decays exponentially in the
number of links.  Unfortunately, technical and fundamental limits on
effectively moving entanglement over even a single link further
complicate the ideal picture and have led to elaborate protocols
involving the distribution, storage, and purification of entangled
states.  The most direct approach is the quantum repeater which has
been proposed to overcome these limitations on a one-dimensional chain
of nodes~\cite{BDCZ98,DBCZ99,CTSL05,HKBD07,SSRG11,VSLMN12}.  There are
examples of practical, deployed quantum networks, such as quantum key
distribution networks. But the technical challenges in
implementing quantum repeaters remain too great to be useful in
contemporary quantum key distribution networks~\cite{SBCD09}. Typically, entanglement is
established over only a single link, while at each node information is
processed classically and re-encoded in a quantum state.

A different approach is to use the entire network, rather than a
linear chain, to distribute entanglement. The availability of multiple
paths is used to overcome the the inevitable decay of
entanglement. This leads to models that are immediately more
interesting because it is not obvious how to prove which of two
protocols is better, let alone which protocol is optimal. In fact
percolation theory~\cite{SA91,Gri99} has provided powerful tools for
evaluating protocols. The best protocols use quantum operations to
transform the initial lattice into a different
lattice~\cite{ACL07,LWL09,PCL+10}.

As in the one-dimensional case, more realistic studies of
multi-dimensional networks have been done, for instance by considering
mixed states and imperfect quantum
operations~\cite{CC09,BDJ09,BDJ10,LPLA11,CC11}.  But sharp questions,
say in the thermodynamic limit, are difficult to pose in these
dirtier situations because of the decay of entanglement. Furthermore,
questions about asymptotic behavior remain that are not only of
intrinsic interest, but address fundamental limits on entanglement
distribution. These are the questions that we address here.

This paper has two main goals. The first goal is to show that enhanced
entanglement percolation (defined below) via lattice transformation is
possible even if the coordination number of the transformed lattice
decreases or the classical percolation threshold increases. The second
goal is to introduce a new tool that we call 
\introword{partial entanglement swapping}.
 In partial swapping, we simply stop the swapping procedure
after the first step, the projection, and evaluate whether the output
state and the new geometry may be more profitably used in a different
operation.  In fact, the usefulness of the tool is demonstrated by
using it to accomplish the first goal. Although the idea behind
partial swapping is simple, it introduces a complication. In
previous entanglement percolation protocols, the Bell measurement
in the computational basis is optimal. But, the optimal basis
for partial swapping is not obvious and depends on the amount
of initial entanglement.

\section{Entanglement Percolation}

Entanglement percolation is described in detail in several
sources~\cite{ACL07,PCA+08,LWL09,PLCLA13,Per10b}. Here we give only a
brief description. We consider the following class of entanglement
percolation models. Each node consists of a collection of qubits. Each
edge, or link, consists of a partially entangled pure state between
two qubits, each on a different node. These states
$\ket{\alpha}\in\mathbb{C}^2\otimes\mathbb{C}^2$ are written in a Schmidt basis as
\begin{equation*}
    \ket{\alpha}=\sqrt{\alpha_0}\ket{00}+\sqrt{\alpha_1}\ket{11},
\end{equation*}
%    \label{stateAB}
where the Schmidt coefficients $\alpha_0,\alpha_1$ satisfy
$\alpha_0\geq\alpha_1$ and $\alpha_0+\alpha_1=1$.
If $\alpha_0=\alpha_1=1/2$, the state is maximally entangled,
and is called a Bell pair or singlet. If either $\alpha_0$ or 
$\alpha_1$ vanishes, then the state is separable
and is useless for quantum information tasks. The smallest Schmidt coefficient
may be used as a measure of entanglement with the amount of entanglement
increasing with $\alpha_1$.
%Furthermore, we consider here only regular lattices.
\begin{figure}
  \includegraphics[width=1 \columnwidth]{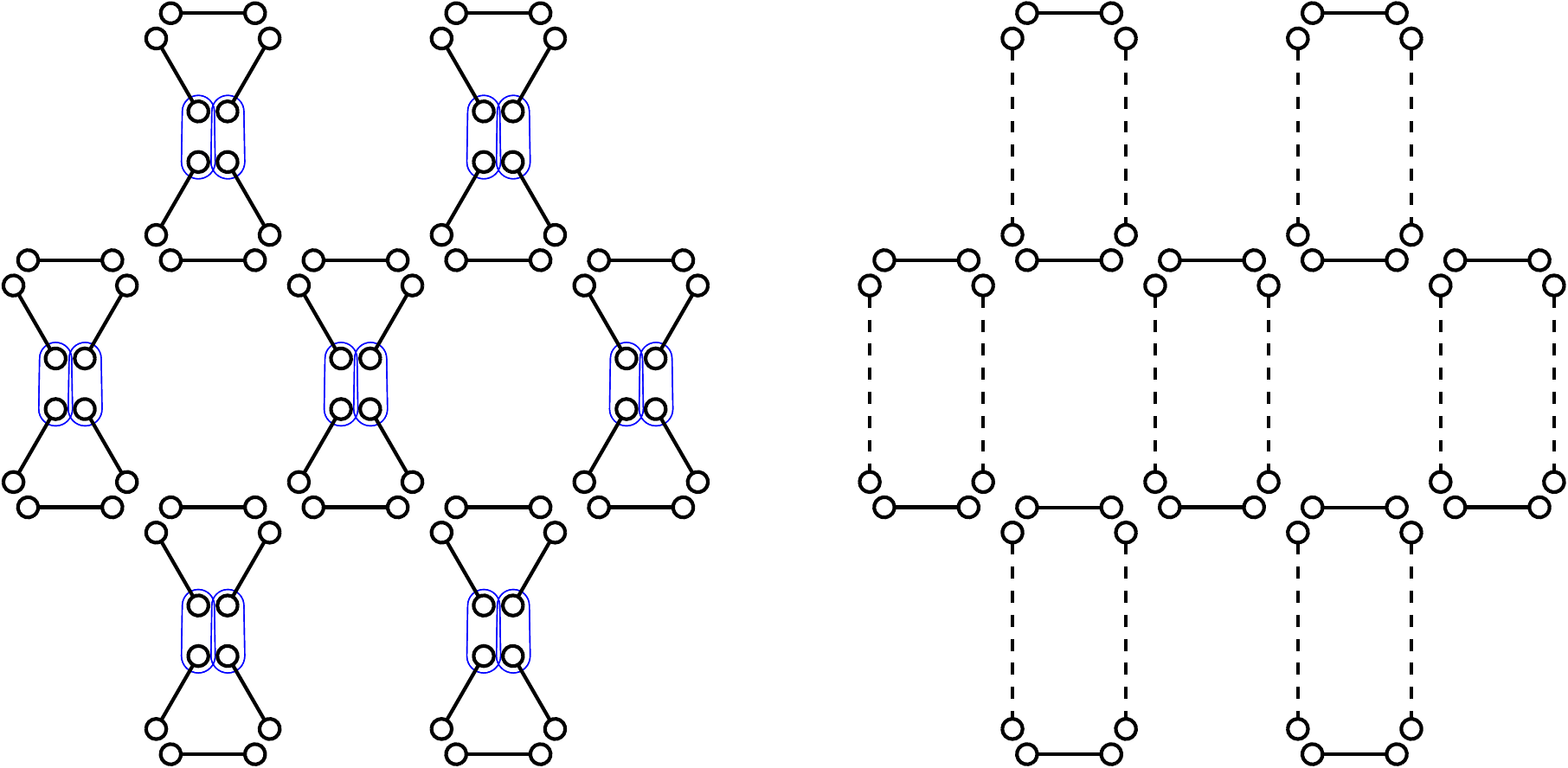}
\caption{Transformation of kagome to square lattice.
Circles represent qubits. Lines represent partially entangled bi-partite states.
\textit{Left}) Full entanglement swapping is performed for each pair of links marked with a
(blue) loop. \textit{Right}) The result is the square lattice, where the vertical (dashed) links
are the outcome of the swap and the horizontal links remain in the state $\initstate$.
The remainder of the QEP protocol is described in the text. }
\label{fig:kagfive}
\end{figure}
The lattice is initialized with identical states $\ket{\alpha}$ on
each link. So we have one free parameter $\alpha_1$.  The goal is to
design a protocol to maximally entangle two arbitrary nodes $A$ and
$B$. The utility of the protocol is measured by the probability of
success $\prAB$ as the distance between $A$ and $B$ tends to
infinity. We require that the protocols use only local operations and
classical communication (LOCC)~\cite{NC00}. This means that quantum
operations that include interaction between qubits on different nodes
are not allowed. But classical communication between all nodes is
allowed.

\mysubsec{Classical Entanglement Percolation} The simplest
entanglement distribution protocol is called 
\introword{classical entanglement percolation} (CEP). 
For some lattices, better protocols have been
found, the so-called \introword{quantum entanglement percolation} (QEP)
protocols. The reason for this distinction and the relation between
CEP and QEP will be made clear below. For now, we note that it is QEP
that uses lattice transformation.  We introduce the CEP and QEP using
the kagome lattice shown in \figref{fig:kagfive}, because it allows a
concise exposition.  First we describe CEP. For the moment, consider
choosing fixed $A$ and $B$. In step $1$ we perform an LOCC operation
on each link, optimally converting it with probability $p=2\alpha_1$ to a Bell
pair, and probability $1-p$ to a separable state. This operation is
called a singlet conversion and $p$ is the singlet conversion
probability (SCP).
After step $1$, we have a lattice in which each link is either open (a
Bell pair), or closed (separable).  In step $2$, we search for an
unbroken path of open links between $A$ and $B$. If no such path
exists, then $\prAB=0$. If a path does exist, then at each
intermediate node we perform an entanglement swapping operation.
Because the input links are singlets, each swap succeeds with
probability $1$, so that $\prAB=1$. We call a protocol that
succeeds with probability $1$ deterministic. This description corresponds
exactly to classical bond percolation, with density of open bonds
$p=2\alpha_1$.
\begin{table}[]
\begin{center}
    \begin{tabular}{@{\extracolsep{10pt}} l l }
    Lattice & $p_c$ for bond percolation \\
    \hline      
    triangular & $2\sin(\pi/18)\approx 0.347$\\
    square     & $1/2$\\
    kagome   & $\approx 0.5244053$ \quad {\text MC estimate } \\
    hexagonal  & $1-2\sin(\pi/18)\approx 0.653$\\
    \end{tabular}
\end{center}
\caption{\label{tab:pcs} $p_c$ for bond percolation on some lattices.
  All critical  densities are exact\cite{Gri99} except for $p_c(\text{kagome})$\cite{ziff1997}. }
\end{table}
The critical bond density for the kagome lattice is
$\pckag\approx 0.52$.  Thus $\prAB=0$ if $p<\pckag$
and $\prAB>0$ if $p>\pckag$.

\mysubsec{Quantum Entanglement Percolation}
A QEP scheme for the kagome lattice is shown in \figref{fig:kagfive}.
We first perform swapping on all pairs of qubits enclosed in loops.
Each of the input states is $\ket{\alpha}$, so the probability of
obtaining a singlet in the resulting vertical link is $p=2\alpha_1$.
We then perform a singlet conversion on the remaining horizontal
bonds, resulting in a square lattice where each link is a Bell pair
with probability $p$ and is separable otherwise.  Finally we perform step
$2$ of CEP (swapping with singlets) on this square lattice. This is
successful precisely when $p>\pcsq$, where $\pcsq$ is the critical
density for bond percolation on the square lattice.  Since
$\pcsq=1/2$, it follows that long-distance entanglement on the kagome
lattice is possible with this QEP scheme, but not with CEP, if
$\alpha_1$ satisfies $\pcsq < 2\alpha_1 < \pckag$.

CEP always gives an easily computable upper bound on the minimum
initial entanglement required for long-distance entanglement. Thus,
CEP serves as a benchmark to compare with any QEP protocol.  Because
we are not interested in QEPs that perform worse than CEP, we will
call any advantageous QEP simply a QEP. However, the measure by which
the QEP is advantageous may vary. We call the smallest value of
$\alpha_1$ such that long-range entanglement is possible the lower
threshold or percolation threshold $\qpc$.  We call the smallest value
of $\alpha_1$ such that long-range entanglement is achieved with
probability $1$ the upper threshold $\qpchi$. Note that $\qpc$ marks a
phase transition, but $\qpchi$ does not. For every lattice, CEP gives
$\qpchi=1/2$.  We call a QEP \introword{robust} if it satisfies at
least one of two conditions.  1) that it lowers the percolation
threshold $\qpc$, and 2) that the upper threshold satisfies
$\qpchi<1/2$ . We are interested in isolating the effect of the
geometry of the transformed lattice on the performance of the QEP. We
therefore emphasize that we will compare the geometry of the
\textit{classical} transformed lattice to that of the the initial
lattice with no reference to quantum states.

\mysubsec{Lattice structure and entanglement distribution}
For any lattice, CEP is defined and the relevant quantities can be taken
directly from percolation theory. But there is no generic prescription
for constructing a QEP. In previous work, QEP protocols have been
identified by choosing a lattice and searching for good lattice
transformations.

In the example above, the initial lattice was transformed into
one new lattice. However, in general, transformations may take
the initial lattice $\initlat$ to
multiple, decoupled lattices $\{{\cal L}'_i\}$~\cite{LWL09}.
It is reasonable to search for $\{{\cal L}'_i\}$ that are
more highly connected than $\initlat$.
In fact, in all of the examples of QEP given in
refs~\cite{ACL07,PCA+08,LWL09,PCL+10}
one ${\cal L}'_i$ has average coordination number greater than
or equal to that of $\initlat$ (condition \textit{i}). Furthermore, one ${\cal L}'_i$ 
has a classical percolation threshold that is less than or equal to
that of $\initlat$ (condition \textit{ii}). In Refs.~\cite{ACL07,PCA+08,LWL09} this is
easy to see because the lattices involved are well-known\footnote{%
These are: double-bond hexagonal to triangular;
square to two square; kagome to square;
bowtie to square and triangular; asymmetric
triangular to square and triangular.}.
The protocols in Ref~\cite{PCL+10} generate multi-partite entanglement
from the initial bi-partite states. The multi-partite swapping was
explicitly designed to increase connectivity. These protocols give the
best performance to date, and often result in less common or
unclassified lattices including non-planar graphs and lattices whose
sites have different coordination numbers.

Given that all known protocols satisfy conditions \condone\ and
\condtwo, a natural question is whether this must always be the
case. Must these properties, one local and one global, that
are associated with high connectivity, be non-decreasing in an advantageous
QEP? In the following section we present a counter-example
demonstrating that the answer to this question is ``no''. In fact both
conditions are violated, and the improvement is robust. To achieve
this, we introduce a new ingredient into the lattice transformation
protocols.

\section{QEP for the triangular lattice}

CEP on the triangular lattice corresponds to classical bond
percolation. With CEP, long-range entanglement is only possible for
$\alpha_1> \qpc(\text{CEP}) = p_c^\triangle/2 \approx 0.1736$, and
deterministic long-range entanglement is only possible for maximally
entangled initial states, \textit{i.e}
$\alpha_1=\qpchi(\text{CEP})=1/2$.
Here we present a QEP that transforms the triangular lattice into the
hexagonal lattice on which a singlet can be created between any two
nodes with probability $1$ if $\alpha_1 \gtrsim 0.3246$. That is,
the upper threshold $\qpchi$ is lowered. This is
possible even though, classically, the hexagonal lattice has larger
critical density $p_c$ and smaller coordination number than the triangular lattice.

\mysubsec{Partial entanglement swapping}
\begin{figure}
 \includegraphics[width=.8 \columnwidth]{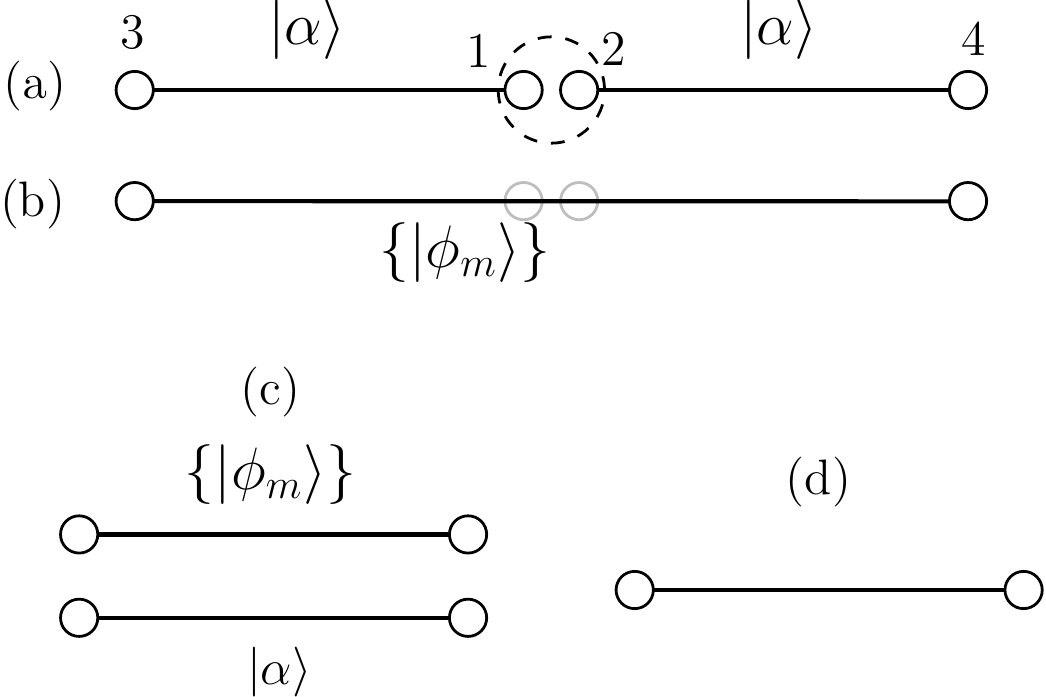}
\caption{Entanglement swapping. (a) a measurement is performed on
qubits $1$ and $2$. (b) after the measurement, qubits
$3$ and $4$ are in one of four states $\{\outcome\}$, which are 
partially or maximally entangled. In full entanglement
swapping, a singlet conversion is performed on the pair $(3,4)$
which results in either a maximally entangled state, or a separable
state. In partial entanglement
swapping, only the measurement is performed. (c) A distillation is
then performed on the output state together with another entangled pair,
here in the state $\initstate$, to produce (d) 
either a more highly entangled pair, or a separable state.}
\label{fig:swap}
\end{figure}
In order to show the counter-example promised in the introduction, it
is enough to consider one of the outcomes from the same swapping
measurement used in previous studies on entanglement percolation.
However, we consider here more general measurements that allow us to
optimize for certain figures of merit. In this paper we consider
entanglement swapping using Bell measurements on two qubits, one from
each pair, as shown in \figref{fig:swap}). For brevity, we omit
referring to any necessary local unitaries. We call the usual
entanglement swapping in any of these bases \introword{full
  entanglement swapping}.  We shall always assume that the two input
pairs are in the same state $\initstate$.  Following
Ref.~\cite{PCA+08}, we define an orthonormal basis
\hbox{$\{\ket{\arrowU},\ket{\arrowD}\}_j$} for each qubit $j=1,2$
\begin{equation*}
    \begin{pmatrix}\ket{\arrowU}\\\ket{\arrowD}\end{pmatrix}_j=U_j
    \begin{pmatrix}\ket{0}\\\ket{1}\end{pmatrix}_j,\quad U_j\in\mathcal{U}(2),
\end{equation*}
and the Bell vectors
\begin{equation*}
    \ket{\Phi^{\pm}}=\frac{\ket{\arrowUU}\pm\ket{\arrowDD}}{\sqrt{2}}
    \quad\text{and}\quad
    \ket{\Psi^{\pm}}=\frac{\ket{\arrowUD}\pm\ket{\arrowDU}}{\sqrt{2}}.
\end{equation*}
The four measurement outcomes are
\begin{equation*}
 \{\outcome\} \quad \text{ with probabilities } 
 \quad \{p_m\}.
\end{equation*}
Furthermore, $\pmin = \min\pset$, and
$\pmax = \max\pset$ are given by
\begin{equation*}
 \pmin = \alpha_0\alpha_1 \quad \text{ and }  \quad
  \pmax = \frac{1}{2} - \alpha_0\alpha_1.
\end{equation*}
There is a bijective mapping between the probabilities $\{p_m\}$
and $\{U_j\}$. In particular,
every (orderless) choice of $\{p_m\}$ satisfying $\pmin\le p_m\le \pmax$
and $\sum_m p_m=1$ corresponds to a Bell measurement.
The smallest Schmidt coefficients of the output
states are given by
\begin{equation*}
 \lambda_m = \frac{1}{2}\left(1-\sqrt{1-\frac{\alpha_0^2\alpha_1^2}{p_m^2}}\right).
\end{equation*}
In full entanglement swapping, we first perform the Bell measurement,
and then perform a singlet conversion on the output state.  Since a
singlet conversion succeeds with probability equal to twice the
smallest Schmidt coefficient, the average SCP for full entanglement
swapping is given by $S_M=2\sum_m p_m\lambda_m$.

However, in partial entanglement swapping, we perform the Bell
measurement only, and not the singlet conversion.  Instead of
immediately doing a singlet conversion we take advantage of the new
geometry of the output state.  We attempt to distill a singlet from
the output state and another entangled pair. Although this is a simple
idea, it is quite useful, and it has not been used in previous work on
entanglement distribution.  From majorization
theory~\cite{Nie99,NV01}, we find that we can distill a Bell pair from
two partially entangled pairs with optimal probability
\begin{equation}
  p_{\text{distill}} = \min\{1,2\left[1-(1-\beta_1)(1-\gamma_1)\right]\},
 \label{pdistill}
\end{equation}
where $\beta_1$ and $\gamma_1$ are the smallest Schmidt coefficients
of the input states~\cite{LWL09}.
In the example below, the second state used in the distillation will
be $\initstate$. Thus, the input states to the distillation have
$\beta_1=\alpha_1$ and $\gamma_1=\lambda_m$.
The average SCP from combining the partial swapping with
distillation is then
\begin{equation}
 S_M = \sum_m p_m \min \left\{1,2-\alpha_0
   \left(1+\sqrt{1-\frac{\alpha_0^2\alpha_1^2}{p_m^2}}\right) \right\}.
 \label{scpgen}
\end{equation}
\subsubsection{Swapping in ZZ basis}
Suppose the measurement is in the ZZ basis, $U_1=U_2=\id_2$.
This is the measurement that maximizes the average SCP in
full swapping. Thus, it is the one used in all previous
entanglement percolation schemes (with a modified version
for multi-partite entanglement percolation).
In this case, $p_1=p_2=\pmin$ and $p_3=p_4=\pmax$, with
corresponding smallest Schmidt coefficients
\begin{equation*}
  \lambda(p_{\max})=\frac{\alpha_1^2}{\alpha_0^2+\alpha_1^2},\quad
  \lambda(p_{\min})=\frac{1}{2}.
\end{equation*}
Two of the outcomes are already singlets. 
Each of the other two may be distilled together with $\initstate$ into a singlet
with probability
\begin{equation}
  p = \min\left\{1,2\left(1-\frac{\alpha^3_0}{\alpha_0^2+\alpha_1^2}\right)\right\},
 \label{zzprob}
\end{equation}
given by \eqref{pdistill}.
The average SCP using partial swapping in the ZZ basis
is then
\begin{equation}
 S_{ZZ} = \alpha_0\alpha_1 +
  (1-2\alpha_0\alpha_1)
 \min\left\{1,2\left(1-\frac{\alpha^3_0}{\alpha_0^2+\alpha_1^2}\right)\right\}.
 \label{zzavg}
\end{equation}
\subsubsection{Swapping in XZ basis}
Suppose the measurement is in the XZ basis. Then
$p_m=1/4$ and $\lambda_m=\frac{1}{2}(1-\sqrt{1-16\alpha_0^2\alpha_1^2})$
for all $m$. The average SCP using partial swapping in the XZ basis
is then
\begin{equation}
 S_{XZ} = \min\left\{1,
 2-\alpha_0\left(1+\sqrt{1-16\alpha_0^2\alpha_1^2}\right)\right\}.
 \label{xzavg}
\end{equation}

\begin{figure}
  \includegraphics[width=.8 \columnwidth]{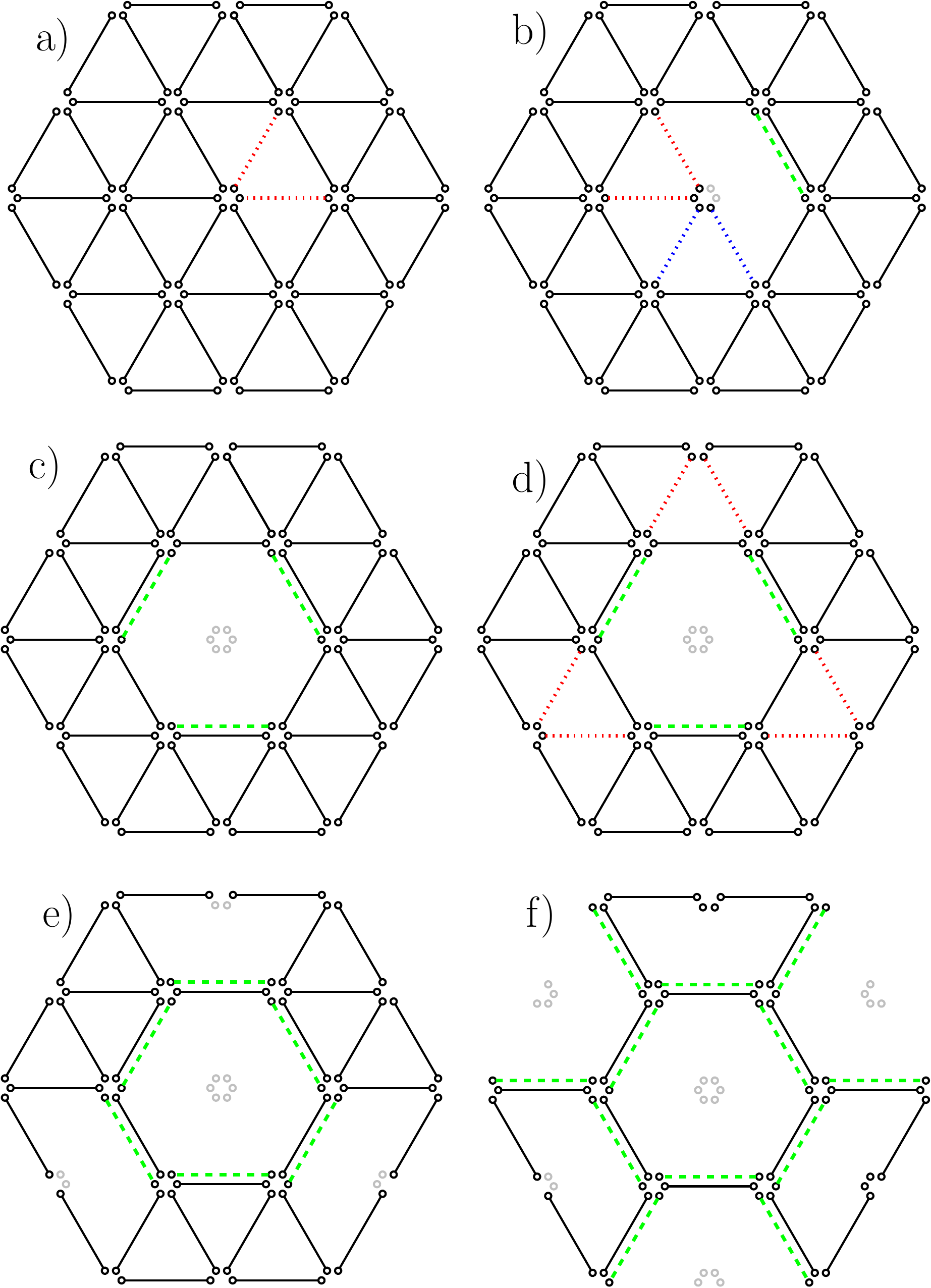}
\caption{Transformation of triangular to hexagonal lattice.
a) triangular lattice. A partial swap is applied to the dotted (red) lines.
In following frames, outcomes of partial swaps are shown as dashed (green)
lines. Partial swaps are applied to pairs of links shown as dotted lines.
f) portion of hexagonal lattice with double links. One link of each double link
is in the state $\initstate$, the other link is one of the four
outcomes of the partial swap $\outcomeset$.}
\label{fig:tritohex}
\end{figure}

\mysubsec{The protocol}
The QEP protocol proceeds as follows.  Consider the triangular lattice
with each bond consisting of a single, partially entangled pure
state. In step $1$, we perform
\textit{partial} entanglement swappings on selected bonds as shown in
\figref{fig:tritohex}.
At the end of step $1$ we have a hexagonal lattice with double links. In each
pair, one link is the initial state $\ket{\alpha}$ and one link is one
of $\{\outcome\}$. Step $2$ consists of the following. For each double
link, if the outcome $\outcome$ is already a Bell pair, then we do
nothing. Otherwise, we attempt to distill a singlet from the two links
$\outcome$ and $\initstate$.
\subsubsection{Protocol in ZZ basis}
Suppose we do the partial swap in the ZZ basis. Two outcomes are
singlets and two are partially entangled. From \eqref{zzprob}
we see that we create a singlet
on every bond of the hexagonal lattice deterministically if
$\alpha_0$ is less than the real root
 $\alpha_0^*\approx 0.6478$ of
 \hbox{$\alpha_0^3-\alpha_0^2+\alpha_0-1/2=0$}. Equivalently, the
condition is $\qpchi \approx 0.3522$.
The critical threshold for this protocol is found by
using \eqref{zzavg} and solving $S_{ZZ}(\alpha_1)=\pchex$,
where $\pchex$ is the classical threshold on the hexagonal lattice,
with the result $\qpc \approx 0.1988$.

\subsubsection{Protocol in XZ basis}
Suppose we do the partial swap in the XZ basis. The smallest
value of $\alpha_1$ for which \eqref{xzavg} equals $1$ is
$\qpchi \approx 0.3246$. The solution of
$S_{XZ}(\alpha_1)=\pchex$ is $\qpc \approx 0.2200$.

\begin{table}[]
\begin{center}
    \begin{tabular}{@{\extracolsep{10pt}} l l l }
    Protocol & $\qpc$ & $\qpchi$ \\
    \hline      
    CEP  & 0.1736 & 1/2 \\
    QEP ZZ  & 0.1988 & 0.3522 \\
    QEP XZ  & 0.2200 & 0.3246 \\
    QEP optimal  & 0.1961 & 0.3246 \\
    \end{tabular}
\end{center}
\caption{\label{tab:pcsqep}
 Percolation thresholds $\qpc$ and 
 upper thresholds $\qpchi$ for entanglement protocols on the
 triangular lattice.}
\end{table}

\begin{figure}
  \includegraphics[width=1.1 \columnwidth]{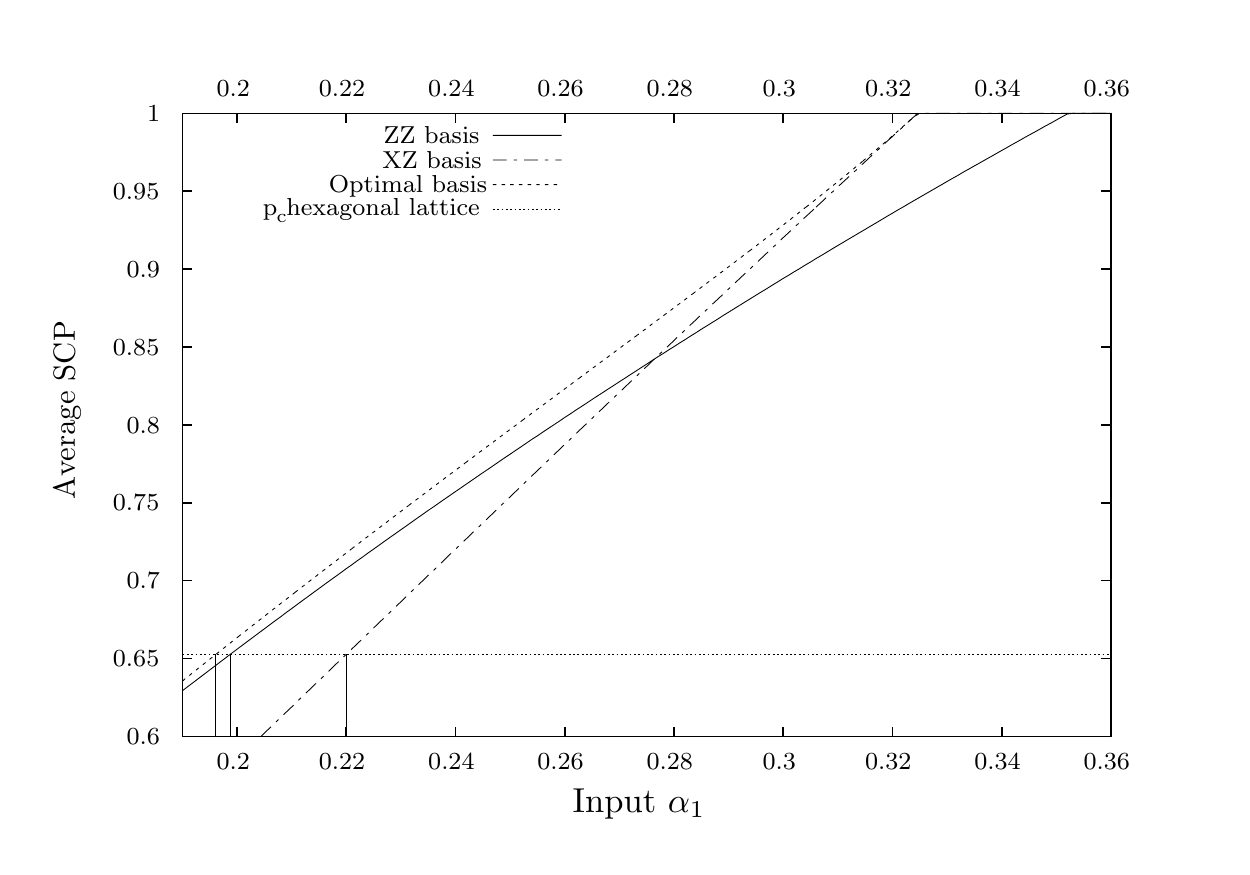}
\caption{Average SCP distilled from double links on the hexagonal
lattice. Solid curve, ZZ basis. Dash-dot curve, XZ basis.
Dashed curve, optimal basis. Dotted line $p_c$ for bond percolation
on the hexagonal lattice.}
\label{fig:scp_opt}
\end{figure}

\subsubsection{Protocol in other Bell bases}
We optimized over all Bell measurements with only two distinct values
of $p_m$. Visual inspection showed that the optimum average SCP occurs
when the second argument to $\min$ in \eqref{xzavg} is equal to $1$,
which occurs for
$p_1=\alpha_0^2\alpha_1/\sqrt{1-2\alpha_1}$. Inserting this into
\eqref{scpgen}, and solving $S_{M}(\alpha_1)=\pchex$,
we find the lower threshold $\qpc \approx 0.1961$, which
is a small improvement over the ZZ basis. Optimizing for the upper
threshold $\qpchi$, we find $p_m=1/4$, which is the XZ basis. A
numerical search for more general Bell measurements strongly suggests
that the optimum Bell protocol has only two distinct values of $p_m$
for all $\alpha_1$.

In summary, we found that the optimal Bell basis has exactly two
distinct values of $p_m$, which depend on $\alpha_1$. At the lower
threshold, the optimal basis gives only a slight improvement over the
ZZ basis. As the upper threshold is approached, the optimal basis
approaches the XZ basis. These results are summarized in
\figref{fig:scp_opt} and Table~\ref{tab:pcsqep}.  We did not
investigate non-Bell measurements.

\section{Discussion}

We have introduced a new tool, partial entanglement swapping, for
entanglement percolation via lattice transformation. This adds
flexibility in optimally combining the quantum and the geometric
aspects of QEPs. We have demonstrated the utility of partial swapping
by using it to design a QEP that transforms the triangular lattice to
the hexagonal lattice. Partial entanglement swapping allows sufficient
concentration of entanglement to overcome lowered connectivity in the
transformed lattice. In particular, there is a least initial amount of
entanglement above which long-distance entanglement is deterministic.
Thus, we have proven that non-decreasing connectivity, as measured by
coordination number and percolation threshold, is not required for
QEP. However, in the present example, we find that CEP still provides
the optimal percolation threshold.  It is interesting to note that the
only other known QEP for the triangular lattice uses multi-partite
entanglement to enhance the connectivity of the lattice by creating a
non-planar graph~\cite{PCL+10}. Thus, the question of whether a
transformed lattice with lower connectivity can give a lower critical
threshold remains open. Also unknown is whether the critical threshold
of the triangular lattice can be lowered via a transformation to a
planar graph, or whether the triangular lattice is, in a sense, a
maximally connected planar graph.

In addition to answering a conjecture on the geometrical constraints
on QEP, partial swapping enlarges the toolbox for QEP. It may be
combined with other techniques to push the initial entanglement
thresholds lower. However, even in this simple example, the search
becomes more complicated because we find that the optimal measurement
basis for partial swapping depends on the amount of initial
entanglement.  Still more interesting than each new protocol would be
a proof, constructive or otherwise, of the existence of a minimum
threshold for a particular lattice or class of lattices.

\section{Acknowledgments}
The author thanks Jan Wehr for discussions and for asking a question
that led to the present work.  This work was supported in part by the
Spanish MICINN (TOQATA, FIS2008-00784), by the ERC (QUAGATUA, OSYRIS),
and EU projects SIQS, EQUAM, and the Templeton Foundation.

\bibliography{tritohex}

\end{document}